\documentclass{jfm}

\usepackage{graphicx}
\usepackage{newtxtext}
\usepackage{newtxmath}
\usepackage{natbib}
\usepackage{hyperref}
\hypersetup{
    colorlinks = true,
    urlcolor   = blue,
    citecolor  = black,
}

\newcommand{\RomanNumeralCaps}[1]

\title{Flow control-oriented coherent mode prediction via Grassmann-kNN manifold learning}

\author{Hongfu Zhang\aff{1},
  Hui Tang\aff{1}
  \corresp{\email{h.tang@polyu.edu.hk}}
 \and Bernd R. Noack\aff{2,} \aff{3}} 
\affiliation{\aff{1}Department of Mechanical Engineering, The Hong Kong Polytechnic University, Hong Kong,
\aff{2}Chair of Artificial Intelligence and Aerodynamics, School of Mechanical Engineering and Automation, Harbin Institute of Technology, 518055 Shenzhen, P. R. China
\aff{3}Guangdong Provincial Key Laboratory of Intelligent Morphing Mechanisms and Adaptive Robotics, Harbin Institute of Technology, Shenzhen 518055, China}

\begin{document}
\maketitle

\begin{abstract}
A data-driven method using Grassmann manifold learning is proposed to identify a low-dimensional actuation manifold for flow-controlled fluid flows. The snapshot flow field are twice compressed using Proper Orthogonal Decomposition (POD) and a diffusion model. Key steps of the actuation manifold are Grassmann manifold-based Polynomial Chaos Expansion (PCE) as the encoder and K-nearest neighbor regression (kNN) as the decoder. This methodology is first tested on a simple dielectric cylinder in a homogeneous electric field to predict the out-of-sample electric field, demonstrating fast and accurate performance. Next, the present model is evaluated by predicting dynamic coherence modes of an oscillating-rotation cylinder. The cylinder's oscillating rotation amplitude and frequency are regarded as independent control parameters. The mean mode and the first dynamic mode are selected as the representative cases to test present model. For the mean mode, the Grassman manifold describes all parameterized modes with 8 latent variables. All the modes can be divided into four clusters, and they share similar features but with different wake length. For the dynamic mode, the Grassman manifold describes all modes with 12 latent variables. All the modes can be divided into three clusters. Intriguingly, each cluster is aligned with clear physical meanings. One describes the near-wake periodic vortex shedding resembling Karman vortices, one describes the far wake periodic vortex shedding, and one shows high-frequency K-H vortices shedding. Moreover, Grassmann-kNN manifold learning can accurately predict the modes. It is possible to estimate the full flow state with small reconstruction errors just by knowing the actuation parameters. This manifold learning model is demonstrated to be crucial for flow control-oriented flow estimation.
\end{abstract}


\section{Introduction}
\label{sec:headings}

Surrogate models have become an important tool for complex dynamic system because they afford a computationally efficient means of approximating (often complex) input–output relations generated from high-fidelity computational models. However, there are very few studies that specifically focus on using surrogate models for parameterized flow control in complex flow dynamics. Most of them focus on a fixed parameter with time-varied snapshot data. In a review paper, \cite{rowley2017model} analyzed the robustness of reduced-order models when applied to both linear and nonlinear problems. These model-based methods aim to extract the main dynamic processes that shape the flow, in order to simplify its modeling. However, the computational benefits obtained from model-based methods come at the cost of simplifying the environment. In other words, the models may not always accurately represent the full dynamics of the system \citep{vignon2023recent}. Therefore, there is a need to develop surrogate models with higher generalization ability. However, this is a significant challenge because flow dynamics with high dimensionality change significantly even with a small shift of a single parameter. 
Surrogate models are often classified as either intrusive or non-intrusive. For non-intrusive methods, \cite{kaiser2014cluster} propose a novel cluster-based reduced-order modeling (CROM) strategy for analyzing unsteady flows. The strategy involves two steps in processing a sequence of time-resolved flow snapshots. First, the snapshot data is clustered into a small number of representative states known as centroids in the state space. These centroids divide the state space into complementary non-overlapping regions called centroidal Voronoi cells. Going beyond the standard algorithm, the probabilities of the clusters are determined, and the states are sorted by analyzing the transition matrix. Second, the transitions between the states are dynamically modeled using a Markov process. The CROM framework is applied to various cases such as the Lorenz attractor, velocity fields of a spatially evolving incompressible mixing layer, and the three-dimensional turbulent wake of a bluff body. Building upon CROM, the authors later develop a ROM for successive bifurcations of a fluidic pinball \citep{deng2021galerkin, deng2022cluster}. \cite{hesthaven2018non} have developed a non-intrusive reduced basis (RB) method for parametrized steady-state partial differential equations (PDEs). The method involves extracting a reduced basis from a set of high-fidelity solutions using POD and utilizes artificial neural networks (ANNs), specifically multi-layer perceptrons (MLPs), to accurately approximate the coefficients of the reduced model. The offline phase includes an automatic routine to search for the optimal number of neurons and the minimum number of training samples to avoid overfitting. This routine relies on a combination of Latin hypercube sampling (LHS) and the Levenberg-Marquardt (LM) training algorithm. The numerical studies presented in the paper focus on driven cavity viscous flows, where the steady incompressible Navier-Stokes equations are used to model the flows. Both physical and geometric parametrizations are considered in their study. \cite{pawar2019deep} presents a modular deep neural network (DNN) framework for data-driven reduced order modeling of dynamical systems relevant to fluid flows. They propose multiple DNN architectures that predict the evolution of dynamical systems by learning from discrete state or slope information. On the other hand, \cite{cheng2020data} introduce an Artificial Intelligence (AI) fluid model based on a general deep convolutional generative adversarial network (DCGAN) to predict spatio-temporal flow distributions. By employing deep convolutional networks, the high-dimensional flows can be transformed into lower-dimensional "latent" representations, allowing for the capture of complex features in flow dynamics through adversarial networks. The DCGAN fluid model mentioned above provides reasonable predictive accuracy for flow fields while maintaining high computational efficiency. \cite{rojas2021reduced} employed proper orthogonal decomposition (POD) to reduce the dimensionality of the model and introduced a novel generative neural ODE (NODE) architecture for predicting the temporal coefficients' behavior. They replaced the traditional Galerkin projection with an architecture that utilizes a continuous latent space. To demonstrate the methodology, they applied it to study the dynamics of the Von Karman vortex street generated by the flow past a cylinder, using a Large-eddy Simulation (LES)-based code. \cite{callaham2022role} investigates nonlinear dimensionality reduction as a means of improving the accuracy and stability of reduced-order models of advection-dominated flows. They used nonlinear correlations between temporal POD coefficients to identify latent low-dimensional structure, approximating the attractor with a minimal set of driving modes and a manifold equation for the remaining modes. They demonstrate this perspective on a quasiperiodic shear-driven cavity flow and show that the dynamics evolves on a torus generated by two independent Stuart–Landau oscillators. The specific approach to nonlinear correlations analysis used in this work is applicable to periodic and quasiperiodic flows and cannot be applied to chaotic or turbulent flows. 
The pure data-driven non-intrusive surrogate models are always lack of generalization. Intrusive surrogate models, which embedded with physical law, could largely improve generalization problem. Intrusive surrogate models, such as those based on Galerkin-POD, have been widely used in the field of laminar fluid dynamics. These models have been successfully applied in numerous studies and the methodology is well-established. More recently, \cite{hijazi2020data} proposed a mixed approach where they analyzed eddy viscosity closure models and incorporated them into a RANS-based ROM. This approach was able to address the solution reproduction problem, and the proposed ROM was validated on benchmark test cases with high Reynolds numbers. In a later study, \cite{hijazi2023pod} introduced the concept of POD-Galerkin Physics-Informed Neural Network (PINN) ROMs. These ROMs utilize deep neural networks to approximate the reduced outputs using the inputs of time and/or model parameters. The effectiveness of this approach was demonstrated through three case studies: steady flow around a backward step, flow around a circular cylinder, and unsteady turbulent flow around a surface-mounted cubic obstacle. However, there have been a few studies that have focused on addressing the parametric problem, where the invariant spaces obtained from POD may vary with different parameters. \cite{lambert2018aerodynamic} proposed a POD-hGreedy ROM, which aims to construct an approximation space that is valid over a range of parameters. They demonstrated the effectiveness of this approach by studying the dependence of a two-dimensional turbulent lid-driven cavity flow on the Reynolds number. However, they found that the ROM was sensitive to changes in the Reynolds number, which may be attributed to the high dimensionality of the Navier-Stokes equation at high Reynolds numbers. In general, intrusive surrogate models provide good convergence in terms of time evolution. However, their complexity limits their flexibility when it comes to problems involving varying parameters. This highlights the need for the development of new numerical schemes that can overcome these limitations and provide more accurate and flexible solutions. Further exploration and research in this area are necessary to address the challenges associated with parameter-dependent surrogate models.  
However, the ROM mentioned above is only applicable to flow dynamics that vary with a single parameter. As the number of parameters increases, their framework becomes more complex and may fail to make accurate predictions. As we indicated above, the key step is to build the map between invariant subspace (POD mode) and parameters. Then, we can project the N-S equation on invariant subspace, forming ODE equations, and solve this simple ODE problem. Grassmannian manifold learning offers several advantages, particularly in handling high-dimensional datasets generated by complex models. One key benefit is its ability to automatically extract important low-dimensional descriptors that effectively capture the complex physics of the system. Furthermore, this approach is well-suited for scenarios with limited data, as it excels in the small-data regime. Additionally, Grassmannian manifold learning significantly reduces training time, minimizes the need for extensive model simulations, and provides a straightforward means of decoding compressed data back to the original space, thereby establishing a link between input parameters and model outputs. Extensive studies have demonstrated its robustness and its ability to accelerate uncertainty quantification tasks in non-linear complex applications \citep{kontolati2022manifold}. Therefore, in this study, we develop Grassmann manifold learning to build the map between invariant subspace (POD mode) and parameters. The illustration of the present method is shown in Figure \ref{tab:k1}
\begin{figure}
  \centerline{\includegraphics[scale=0.5,angle=0]{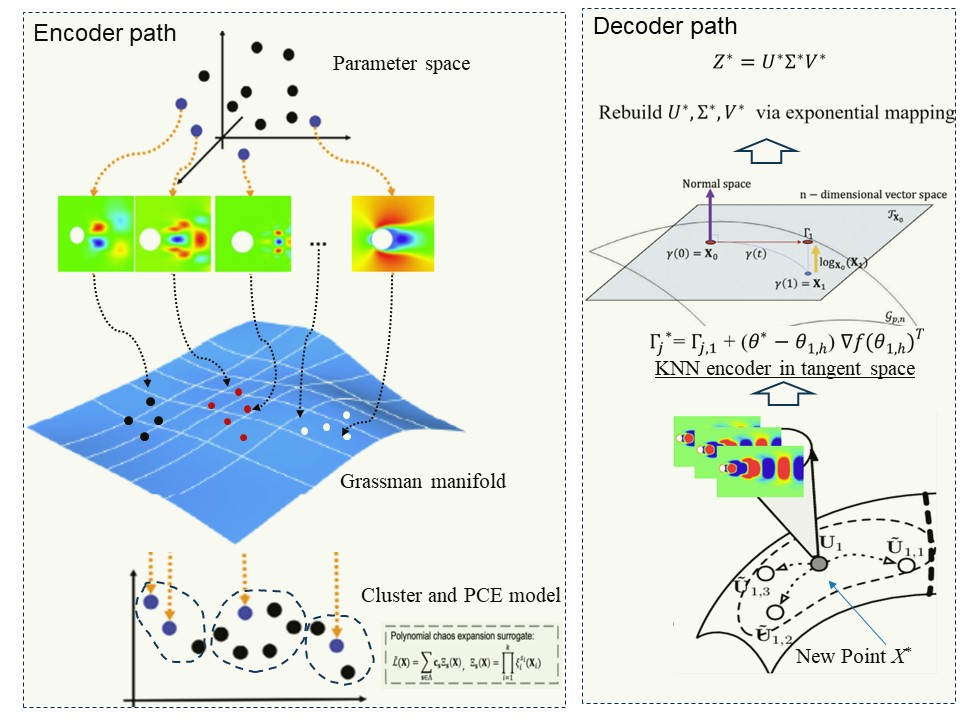}}
  \caption{Illustration of the present method}
\label{fig:ka}
\end{figure}
\section{Grassman-kNN manifold learning}
\label{sec:headings}
Grassmann manifolds allow for a variety of mathematical processes, including differentiation and optimization, due to their smooth and differentiable nature. This property enables them to effectively identify key patterns in intricate data, thereby simplifying complexity and enhancing performance. An element on a Grassmann manifold is typically represented an arbitrarily chosen $n × k$ orthonormal matrix $X$ whose column spans the corresponding subspace, called a generator of the element. A Grassmannian manifold consists of a set of generator matrices that can be expressed mathematically as
\begin{equation}\mathcal{G}(p,n)=\{span(X):X\in\mathbb{R}^{n\times p},\boldsymbol{X}^{\intercal}\boldsymbol{X}=\boldsymbol{I_k}\}\label{e1}\end{equation}

The approach consists of two primary steps: First, the encoder compresses high-dimensional model data onto a low-dimensional Grassmann diffusion manifold, using a Polynomial Chaos Expansion (PCE) surrogate to link the input space to the latent space. Second, the decoder reconstructs the full high-dimensional solutions from the predicted low-dimensional data. The approach is illustrated graphically in Figure \ref{fig:ka}. 
(1) Encoder : The encoder path to produce an inexpensive mapping from the input space to a low-dimensional latent space representation of the high-dimensional response is detailed herein and the corresponding algorithm is provided in Algorithm 1.
Consider parameters of a dynamic system  $X=\{\mathrm{X}_1,..,\mathrm{X}_\mathrm{N}\}\mathrm{~with~}N\mathrm{~random}$ samples $\mathrm{X_i\in\mathbb{R}^k}$ drawn from the joint PDF $\mathrm{Q_X}.$ The dynamic system $\mathcal{M}:\mathrm{X_i}\in\mathbb{R}^\mathrm{k}\to\mathrm{Z_i}\in\mathbb{R}^\mathrm{n\times m}$  is adopted to produce the corresponding dynamic response $\mathrm{Z}=\{\mathrm{Z}_1,\ldots,\mathrm{Z}_\mathrm{N}\}.$ Here, the response $Z$ has high dimensionality, for instance, the grid numbers of the computational domain for fluid dynamic system. It is assumed that the data has already undergone a train-test split to assess the surrogate model’s performance on unseen samples. And, $N$ represents the number of training samples which, for most real-world engineering applications, is rather small.
Given a set of $N$ high-dimensional data points $\mathrm{Z}=\left\{\mathrm{Z}_{1},..,\mathrm{Z}_{\mathrm{N}}\right\}$, each data point $\mathbf{z}_{i}$ is projected onto the Grassmann manifold using a thin singular value decomposition (SVD). 
\begin{equation}\mathrm{Z}_i=U_i{\Sigma_iV_i^T}\label{e2}\end{equation}
where the columns of the matrices  $\mathbf{U}_i\in\mathbb{R}^{n\times p}\mathrm{~and~}\mathbf{V}_i\in\mathbb{R}^{n\times p}$ are orthonormal vectors and they follow $\mathbf{U}_i^\mathrm{T}\mathbf{U}_i=\mathbf{I}_p\mathrm{~and~}\mathbf{V}_i^\mathrm{T}\mathbf{V}_i=\mathbf{I}_p$. $\mathbf{\Sigma}_\mathrm{i}\in\mathbb{R}^{p\times p}$ is a diagonal matrix composed of singular values arranged by magnitude. Therefore, $\mathbf{U}_i, \mathbf{V}_i$ live on the Grassmannians $\mathcal{G}(\mathrm{p},\mathrm{n})=\{\mathrm{span}(\mathrm{U}):\mathrm{U}\in \mathbb{R}^{\mathrm{n\times p}}\}$ and $\mathcal{G}(\mathrm{p},\mathrm{n})=\{\mathrm{span}(\mathrm{V}):\mathrm{V}\in\mathbb{R}^{\mathrm{m\times p}}\}$.
 respectively. Dimension $p$ can be defined a priori or calculated automatically using SVD tolerances.
For each pair of $\begin{bmatrix}\mathrm{U_i,U_j}\end{bmatrix}$ and $\begin{bmatrix}\mathrm{V_i,V_j}\end{bmatrix}$, the corresponding entries in the kernel matrices $\mathrm{k_{ij}(U)}$ and $\mathrm{k_{ij}(V)}$ are computed.
The Binet-Cauchy kernel here is selected, and the mappings are defined as $\mathrm{k_{ij}(U)}$:$\mathcal{G}(\mathrm{p},\mathrm{n})\times \mathcal{G}(\mathrm{p},\mathrm{n})\to\mathbb{R}$ and $\mathrm{k_{ij}(V)}$:$\mathcal{G}(\mathrm{p},\mathrm{m})\times \mathcal{G}(\mathrm{p},\mathrm{m})\to\mathbb{R}$, respectively. The kernel $\mathrm{k_{ij}(U)}$ and $\mathrm{k_{ij}(V)}$  are then used to construct the diagonal matrix $\mathbf{D}\in\mathbb{R}^{N\times N}$ and then the normalized matrix $\kappa$ with components $\kappa_{ij}$ in Eq. \ref{A10}. Next we construct the transition probability matrix $\mathbf{p}^{t}$ of the Markov chain over the data and we perform an eigen-decomposition of $\mathbf{p}^{t}$ to determine the truncated diffusion map basis of ${\boldsymbol{q}}$ eigenvectors $\{\xi_{k}\}_{k=1}^{q}$, with $\left(\xi_{k}\right)\in\mathbb{R}^{N}$
 and corresponding eigenvalues $\{\lambda_k\}_{k=1}^q$. The diffusion coordinates are therefore given by $\Theta=\{\mathbf{\Theta}_1,\ldots,\mathbf{\Theta}_N\}\mathrm{~where~}\Theta_i\in\mathbb{R}^q$. Herein, we will refer to the mapping of the input parameters X to the diffusion coordinates $\mathrm{\Theta},$ as$\mathcal{L}\colon\mathbf{X}_i\in\mathbb{R}^k\to\mathbf{\Theta}_i\in\mathbb{R}^q$. 
 The dimension $\begin{array}{c}q\end{array}$ of the diffusion coordinates $\Theta_i\in\mathbb{R}^q$(embedding) is much smaller than the dimension of the data on the ambient space $\mathbf{Y}_i\in\mathbb{R}^{n\times m}$ ($\mathrm{i.e.},q\ll $n×m) and therefore GDMaps allows us to achieve a significant dimension reduction.
Given a training dataset of input random variable realizations X=$\{\mathrm X_1,\ldots,\mathrm X_{\mathrm N}\},\mathrm X_{\mathrm i}\in\mathbb{R}^{\mathrm k}$, and corresponding solutions projected on the latent space $\mathrm{\Theta}$=$\{\mathbf{\Theta}_1,..,\mathbf{\Theta}_N\},\mathbf{\Theta}_i\in\mathbb{R}^q$, we construct a PCE as explained in Appendix \ref{appA} to approximate the true encoder $\mathcal{L}{:} \mathbf{X}\to\mathbf{\Theta}$ as
\begin{equation}\tilde{\mathcal{L}}(\mathbf{X})=\sum_{s\in\Lambda}c_s\Xi_s(\mathbf{X})\label{e3}\end{equation}
where $\Lambda$ is a total-degree multi-index set, $\Xi_s$ are the multivariate orthonormal polynomials, and the PCE coefficients are now vector-valued with dimension equal to the one of the diffusion coordinates, $\mathrm{i.e.c_s}\in\mathbb{R}^q.$
\\To assess the predictive ability of the PCE surrogate, we employ an error metric known as the generalization error, which is defined as
\begin{equation}\epsilon_{gen}=\mathbb{E}_{\mathbf{X}}\left[\left(\mathcal{L}(\mathbf{X})-\tilde{\mathcal{L}}(\mathbf{X})\right)^2\right]\label{e4}\end{equation}
We approximate $\epsilon_{gen}$ with the validation error, which is computed on a validation dataset of $\mathcal{N}_{*}$ test realizations. The validation error is computed as
\begin{equation}\epsilon_{val}=\frac{\sum_{i=1}^{N_{*}}\left(\mathbf{0}_{i}^{*}-\tilde{\mathcal{L}}(\mathbf{X}_{i}^{*})\right)^{2}}{\sum_{i=1}^{N_{*}}\left(\mathbf{0}_{i}^{*}-\overline{\mathbf{0}^{*}}\right)^{2}}\label{e5}\end{equation}
where $\{\mathbf{X}_{i}^{*}\in\mathbb{R}^{k}\}_{i=1}^{\mathcal{N}_{*}},\{\mathbf{\Theta}_{i}^{*}\in\mathbb{R}^{k}\}_{i=1}^{\mathcal{N}_{*}}\ \mathrm{and}\ \overline{\mathbf{\Theta}^{*}}=\frac{1}{\mathcal{N}_{*}}\sum_{i=1}^{\mathcal{N}_{*}}\mathbf{\Theta}_{i}^{*}$ \ is the mean response. Accordingly, the total degree $s_{max}$ is chosen so that the validation error is minimized. In cases where a validation dataset cannot be generated due to computational constraints, alternative measures such as the $k$-fold cross validation can be considered . However, such techniques introduce different computational costs, as they require the construction of multiple surrogates for different partitioning of training dataset, therefore a prior evaluation of the trade-off and respective costs is required.
K-means algorithm is applied to cluster the diffusion coordinates $\theta_i\in R^q,$ such that the distance between points belonging to one cluster is minimized on the diffusion manifold. We begin with the smallest possible number of clusters, $\mathrm{i.e.},\ell=2.$ At each iteration and for each cluster $C_h\mathrm{~where~}h=2,..,\ell,$ we compute the Karcher means $\mathbf{m}_{u,h},\mathbf{m}_{v,h}$ ,~of~points $\{\mathbf{U}_{i}\}_{i=1}^{N_{h}}\in C_{h}$ and $\{\mathbf{V}_{i}\}_{i=1}^{N_{h}}\in C_{h}$, respectively, where $\mathcal{N}_{h}$ represents the total number of points for a given cluster. For the computation of the Karcher means and the minimization of the loss function we use stochastic gradient descent. 
\begin{table}
\begin{tabular}{ll}
Algorithm 1: Modeling Grassman diffusion manifolds (encoder path)                                                                                               &                         \\
Input: parameters $X=\{X_{1},\ldots,X_{N}\}$ where $X_i\in R^k$, corresponding response variables $Z=\{Z_1,..,Z_N\}$  \\
where $Z_{i}\in R^{n\times m}$ via physical law $Z=\mathcal{F}(X)$                                                                                                                                         &                         \\
Output: Grassman diffusion manifolds-based surrogate model                                                                                                      &                         \\
1. Perform SVD $Z_i=U_i\Sigma_iV_i^T\mathrm{where~}U_i\in G(\mathrm{p},n),V_i\in G(\mathrm{p},m)$                                                                             &                         \\
2. Construct a Grassmannian diffusion kernel $k(Z_{i}),\mathrm{e.g.},$ the Binet-Cauchy kernel                                                                            &                         \\
3. Compute the diffusion coordinates $\{\theta_{i}\in R^{q}\}_{i=1}^{N},$ where $q\ll n\times m$                                                     &                         \\
4. Build a surrogate model via PCE approximation  $\tilde{L}(X)=\Theta(X)=\sum c_s\Xi_s(X)$                                                                                                              &                         \\
5. Perform k-means clustering to identify clusters $\{C_h\}_{h=1}^{\ell}$ with $N_{h}$ points, where $h=1,...,\ell $
                                                             &              \\    
6. Compute the Karcher means$m_{Z,h}$ of points $\{Z_{i}\}_{i=1}^{N_{h}}\in C_{h}$
\end{tabular}
\end{table}
\\(2) Decoder: The predictions of PCE in the low-dimensional latent space (Grassmann diffusion manifold) are propagated to obtain an approximate high-dimensional solution via the decoder path discussed here after and described through Algorithm 2. Give new parameters $X^{*},$ the diffusion manifold coordinate $\mathbf{\Theta}^{*}$ can be obtained by the PCE surrogate. Secondly, we calculate the Euclidean distance between $\mathbf{\Theta}^{*}$ and eatch Karcher means $m_{z,h},$ and clarify the cluster $C_{h}$ in which the unknown point $Y^{*}$ belongs to.  Identify the K points nearest the $\mathbf{\Theta}^{*}$ using KNN in the cluster  $C_{h}$, namely, $\theta_{1,h}$, $\theta_{1,h}$,…$\theta_{K,h}$. Having as origin $[U_{1,h}\Sigma_{1,h},U_{1,h}],$ we project the K points onto to the corresponding tangent spaces $\Gamma_{U,i},\Gamma_{V,i},\Gamma_{\Sigma,i}$ via the logarithmic mapping in Equation \ref{A5}.
In order to provide a decoder to create a correspondence between diffusion coordinates $\mathbf{\Theta}$ and the ones in the tangent space, we employ a purely data-driven approach. Any point in tangent space $[\Gamma_{U},\Gamma_{\Sigma},\Gamma_{V}]\in R^{\mathrm{P}}$ has its low-dimensional counterpart  $\theta\in R^{\mathrm{p}},i=1,...,N.$ To reconstruct the point in tangent space for $\Gamma_{j}^{*}(j=U,\Sigma\mathrm{or}V),$ we assume that its $k$-nearest neighbors $\theta_h=[\theta_{1,h},\theta_{1,h},...\theta_{K,h}]$ and their high-dimensional counterparts, namely $\Gamma_{j,h}=\begin{bmatrix}\Gamma_{j,1},...,\Gamma_{j,K}\end{bmatrix}$, are identified. Therefore, the reconstruction of  $\Gamma_{U,h}$ can be obtained as a first-order Taylor expansion starting from the nearest neighbor to be mapped back to the original space, as
\begin{equation}\Gamma_j^*=\Gamma_{j,1}+\left(\theta^*-\theta_{1,h}\right)\nabla f(\theta_{1,h})^T\end{equation}
where the gradient tensor in $\theta_{1,h},$ namely $\nabla f(\theta_{1,h}),$ is estimated assuming an orthogonal projection of the $K-1$ directions provided by the K-nearest neighbors. This is
\begin{equation}\nabla f(\theta_{1,h})^T=(\Delta\theta_h^T\Delta\theta_h)^{-1}\Delta\theta_h^T\Delta\Gamma_{j,h}\label{e6}\end{equation}
which yields  $\nabla f(\theta_{1,h})^{T}=(\Delta\theta_{h}^{T}\Delta\theta_{h})^{-1}\Delta\theta_{h}^{T}\Delta\Gamma_{j}$ if least squares minimization is used to approximate it, and $\Delta\theta_h\mathrm{~and~}\Delta\Gamma_U$ are the left-hand side and the first term in the right-hand side of (2.4), respectively. 
\begin{equation}\begin{bmatrix}\Gamma_{j,2}&-&\Gamma_{j,1}\\\vdots&\cdots&\vdots\\\Gamma_{j,K}&-&\Gamma_{j,1}\end{bmatrix}\simeq\begin{bmatrix}\theta_{2,h}&-&\theta_{1,h}\\\vdots&\cdots&\vdots\\\theta_{K,h}&-&\theta_{1,h}\end{bmatrix}\nabla f(\theta_{1,h})^T\label{e7}\end{equation}
At last, we calculate $U^*,\Sigma^*,V^*$  via exponential mapping. The folding variable can be computed by$Z^{*}=U^{*}\Sigma^{*}V^{*}$ and physical variable can be obtained by stretching $Y^*\in R^{M\times1}\leftarrow Z^*\in R^{n\times m}.$
 \begin{table}
\begin{tabular}{l}
Algorithm 2: From parameters to physical variables (decoder path)                                                                                                                                 \\
Input: new parameters $X^{*}$                                                                                                                                                        \\
Output: Dynamic system response $Y^{*}$ for testing samples $X^{*}$                                                                                                     \\
1. Compute the diffusion coordinates $\theta^{*}$ with PCE surrogate model                                                                                                                \\
2. Calculate the Euclidean distance between $\theta^{*}$ and eatch Karcher means $m_{Z,h}$, and 
clarify the cluster
 \\
$C_{h}$ in which the unknown point $Y^{*}$ belongs to.                                                                                                                                                                                                     \\
3. Identify the K points nearest the using KNN in the $\theta^{*}$ cluster $C_{h}$,namely$,\theta_{1,h},$ $\theta_{1,h},...\theta_{K,h}$                                                                                                                                                                                                                                                                                         \\
4. Having as origin $[U_{1,h}\Sigma_{1,h},U_{1,h}]$ project the K points onto tangent spaces via  logarithmic mapping, 
 \\
namely, $\Gamma_{U,i},\Gamma_{V,i},\Gamma_{\Sigma,i}(i=2...K)$                                                                                                                                                                    \\
5. Estimate gradient tensor for Grassman points $\Gamma_{U}$ using K$\text{K}$ nearest neighbours  
 \\
$\nabla f\left(\theta_{1,h}\right)^{T}=(\Delta\theta_{h}{}^{T}\Delta\theta_{h})^{-1}\Delta\theta_{h}{}^{T}\Delta\Gamma_{U}.$ Same for $\Gamma_{\Sigma}$ and $\Gamma_{U}$.                                                                                                                                                             \\
6. Compute the target $\Gamma_{U}^{*}=\Gamma_{U,1}+(\theta^{*}-\theta_{1,h})\nabla f(\theta_{1,h})^{T},$  Same for $\Gamma_{\Sigma}^*$ and $\Gamma_V^{*}$ \\
7. Rebuild $U^{*}\Sigma^{*}V^{*}$  via exponential mapping                                                                                                                       \\
8. Compute the folding variable $Z^{*}=U^{*}\Sigma^{*}V^{*}$
\end{tabular}
\end{table}

\section{Results and discussion}\label{sec:Figures_Tables}
\subsection{Dielectric Cylinder in Homogeneous Electric Field}
 In this scenario, we examine a theoretical model derived from electromagnetic field system, specifically focusing on an infinitely extended dielectric cylinder positioned within a uniform electric field (see Figure \ref{fig:k2-1}). The issue can be simplified to a two-dimensional analysis because of the translational symmetry along the $z$-axis. Thus, we set the computational domain as $\Omega=[-1,1]\times[-1,1]$ and the cylinder’s domain by $Dc=\{X=(x,y)|\sqrt{x^2+y^2}\leq r_0\},$ where $r_{0}$ is the cylinder’s radius. 
 The relative permittivity of the dielectric  \( \mathbf{\varepsilon_{\mathrm{c}}}\) represents the permittivity inside the cylinder while \(\mathbf{\varepsilon_{\mathrm{o}}}\) represents the permittivity outside the domain of cylinder. A homogeneous electric field $e_{\infty}=(E_{\infty},0). $ We also consider that the rectangular domain $\Omega$ is equipped with Dirichlet boundary conditions (BCs) on the left and right boundaries and Neumann BCs on the top and bottom of this domain. Using the Laplace equation, it can be shown that $u(x)$ in $\Omega$ is given by
 \begin{equation}-\nabla\cdot\left(\varepsilon(\mathbf{x})\nabla u(\mathbf{x})\right)=0\label{e8}\end{equation}
 \begin{equation}u(\mathbf{x})=u^*(\mathbf{x})\label{e9}\end{equation}
 \begin{equation}\begin{pmatrix}\nabla u(\mathbf{x})\end{pmatrix}\cdot\mathbf{n}=\begin{pmatrix}\nabla u^*(\mathbf{x})\end{pmatrix}\cdot\mathbf{n}\label{e10}\end{equation}
 where n denotes the outer normal unit vector, the permittivity $\varepsilon\left(x\right)$ is given as
 \begin{equation}\varepsilon(\mathbf{x})=\begin{cases}\varepsilon_\mathrm{c},&\mathbf{x}\in D_\mathrm{c}\\\varepsilon_\mathrm{o},&\mathbf{x}\in\Omega\setminus D_\mathrm{c}\end{cases}\label{e11}\end{equation}
 and $u^{*},$ which is also the analytical solution to the problem, is given by
\begin{equation}u^*(\mathbf{x})=-E_\infty x\begin{cases}1-\frac{\frac{\varepsilon_\mathrm{c}}{\varepsilon_0}-1}{\frac{\varepsilon_\mathrm{c}}{\varepsilon_0}+1}\frac{r_0^2}{x^2+y^2},&\mathbf{x}\in\Omega\setminus D_\mathrm{c}\\\frac{2}{\frac{\varepsilon_\mathrm{c}}{\varepsilon_0}+1},&\mathbf{x}\in D_\mathrm{c}\end{cases}\label{e12}\end{equation}

We account for fluctuations in the electric potential due to stochasticity of two input settings $r_{0} = Uniform[0.2, 0.6]$, $\mathrm{E}_{\infty} = Uniform[8,16]$.
Specifically, we create $N$ = 400 training samples $X\in\mathbf{R}^{\mathrm{N}\times2}$, where each is associated to model outputs $Y\in\mathbf{R}^{\mathrm{N}\times6400}$, that discretize the square computational domain in $(w \times w) = (80  × 80) = 6400$ mesh points. The GDMaps converged to Grassmann manifold dimension $p$ = 30 for all training sets withheld from the optimization, and these correspond to matrices on the Grassmann {$\{U_i,V_i\in G(30,80)\}^{\mathrm{N}_{i=1}}$, respectively. The residuals based on the eigen decomposition of Markov matrix for each dataset $N$ = 400 used to determine the first $q$=4 non-trivial diffusion coordinates given by $\{\theta_1,\theta_2,\theta_3,\theta_4\}$ corresponding to the embedding structure.
Figure \ref{fig:k2} show that Grassmann-kNN divided the data into 12 clusters. The prediction result is shown in Figure \ref{fig:k3}, Grassman-kNN can accurately predict the electric potential field not only for the in-sample case but also for out-of-sample case. The prediction error of out-of-sample case is higher than that of the in-sample case, as is expected. We also compared the results with the counterpart of Kontolati et al (2021), see Table II. Here, we use single core to train and predict the results. The CPU applied is Intel(R) Core(TM) i7-9700K @ 3.60GHz. It shows that the present method is much faster and more accurate than that of \cite{kontolati2022manifold}.
\begin{figure}
  \centerline{\includegraphics[scale=0.15,angle=0]{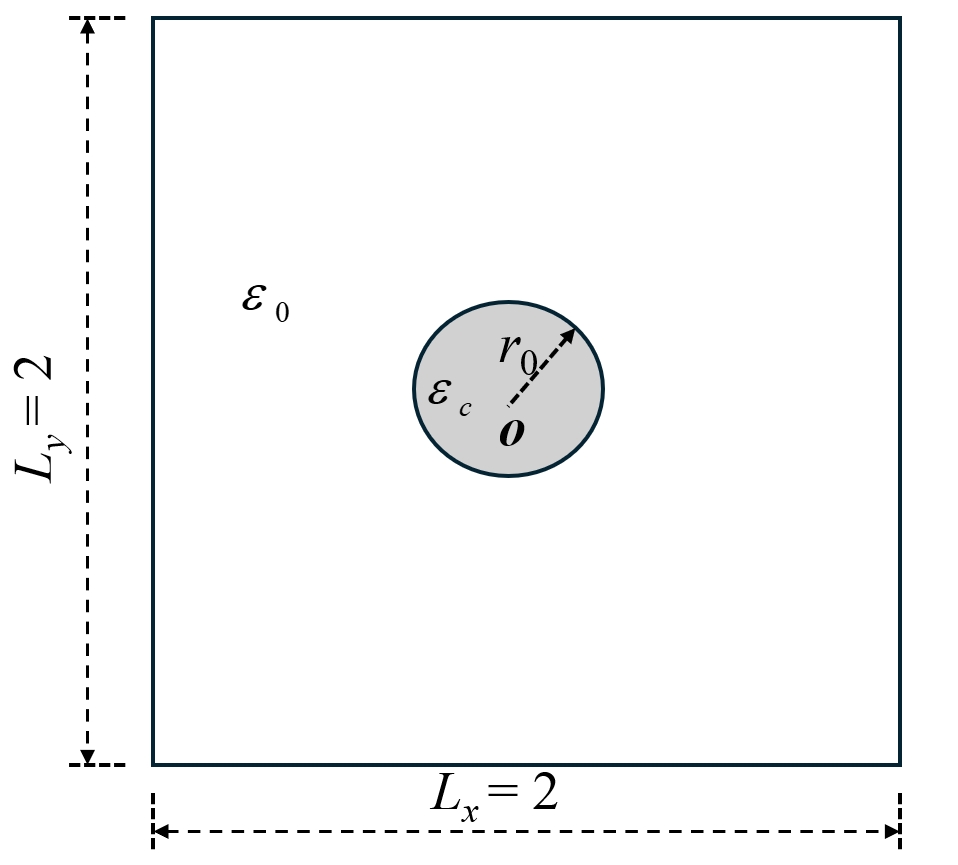}}
  \caption{Dielectric cylinder positioned within a uniform electric field.}
\label{fig:k2-1}
\end{figure}

\begin{figure}
  \centerline{\includegraphics[scale=0.35,angle=0]{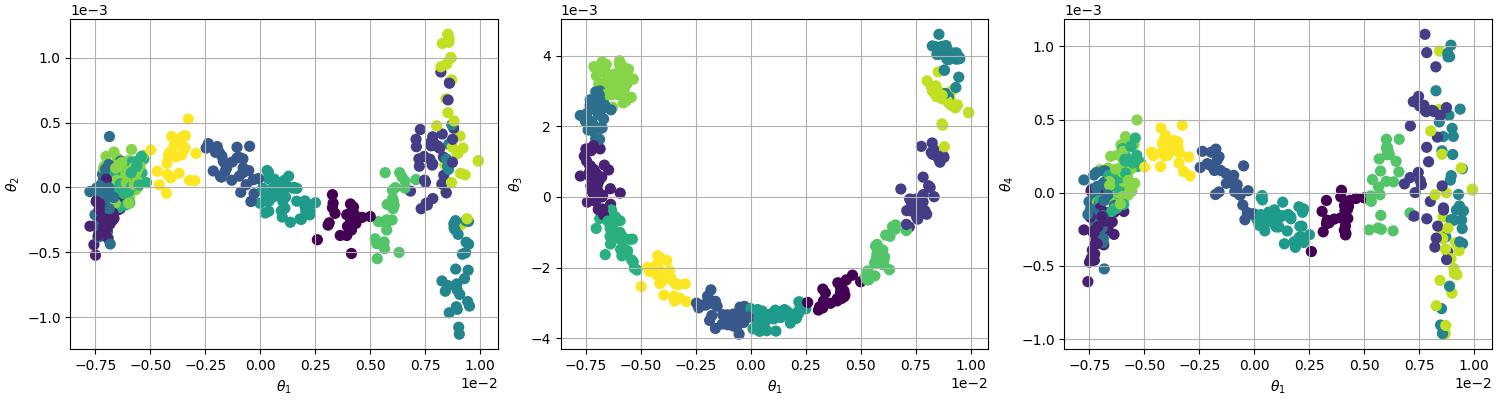}}
  \caption{The diffusion coordinates for Grassmann manifold dimension.}
\label{fig:k2}
\end{figure}
\begin{figure}
  \centerline{\includegraphics[scale=0.2,angle=0]{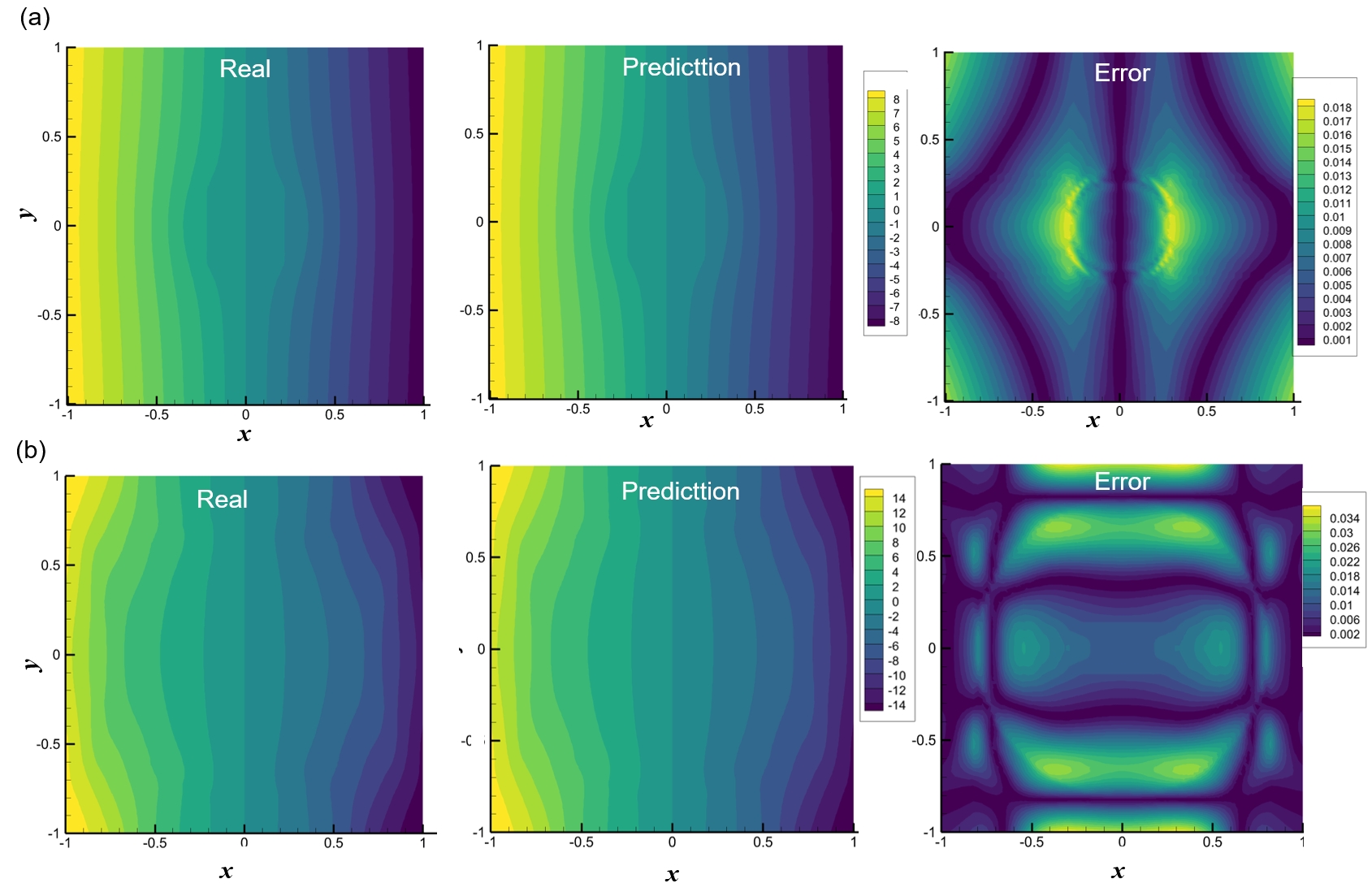}}
  \caption{Reference, prediction, and relative error of the electric potential field. (a) In-sample interpolation with $r_0$ = 0.273, $\mathrm{E}_{\infty}$ = 10.523; (b) Out-of sample  interpolation  $r_0$ = 0.8, $\mathrm{E}_{\infty}$ = 19.0}
\label{fig:k3}
\end{figure}

\begin{table}
  \begin{center}
\def~{\hphantom{0}}
  \begin{tabular}{lcl}
      & RMS error&Time cost\\[3pt]
       Kontolati et al. (2022)& 0.040&502 s\\
       Present result& 0.015&40 s\\
  \end{tabular}
  \caption{Comparation of present results and \cite{kontolati2022manifold}}
  \label{tab:k1}
  \end{center}
\end{table}

\subsection{Oscillating cylinder with self-rotation}
The configuration of the simulated system is illustrated in Figure \ref{fig:k4} The oscillating of the cylinder is chosen to alter the flow pattern around a main cylinder placed upstream within a uniform flow with velocity $U_{\infty}$. In this system, the Reynolds number is defined as $Re=U_{\infty}D/\nu=100,$ where $D$ is the diameter of the main cylinder, and $\nu $ is kinematic viscosity of the fluid. For all cases, the oscillating frequency of the cylinders was set as $\omega=\begin{bmatrix}-6,6\end{bmatrix}$ and the oscillating amplitude set as $A/D=[-2,2].$ The details of the system parameter are shown in Table \ref{tab:k1}.
\\The flow field was obtained from the two-dimensional Navier–Stokes equations: 
\begin{equation}\frac{\partial U}{\partial t}=-(U\cdot\nabla)U-\nabla P+\frac{1}{Re}\nabla^2U+S\label{e13}\end{equation}
\begin{equation}\nabla\cdot U=0\label{e14}\end{equation}
The equations were solved using a high-fidelity solver based on the generalized interpolation-supplemented lattice Boltzmann method (GILBM). To avoid wall effects, the inlet length, sidewall width, and outlet length were set slightly larger than those used by \cite{bhatt2018vibrations} and \cite{bao2012numerical}, as depicted in Figure \ref{fig:k2}.

The free-stream boundary conditions (Dirichlet boundary conditions) are applied at the inlet and the upper/lower side as shown in Fig. \ref{fig:k2}. Outflow boundary conditions of velocity with zero normal gradient and constant zero pressure. The square cylinders also had homogenous Dirichlet boundary conditions for velocity, and a zero normal gradient Neumann boundary condition for pressure. Second-order time accuracy was attained by employing a three-step time-splitting method with an velocity corrector technique.
\begin{figure}
  \centerline{\includegraphics[scale=0.2,angle=0]{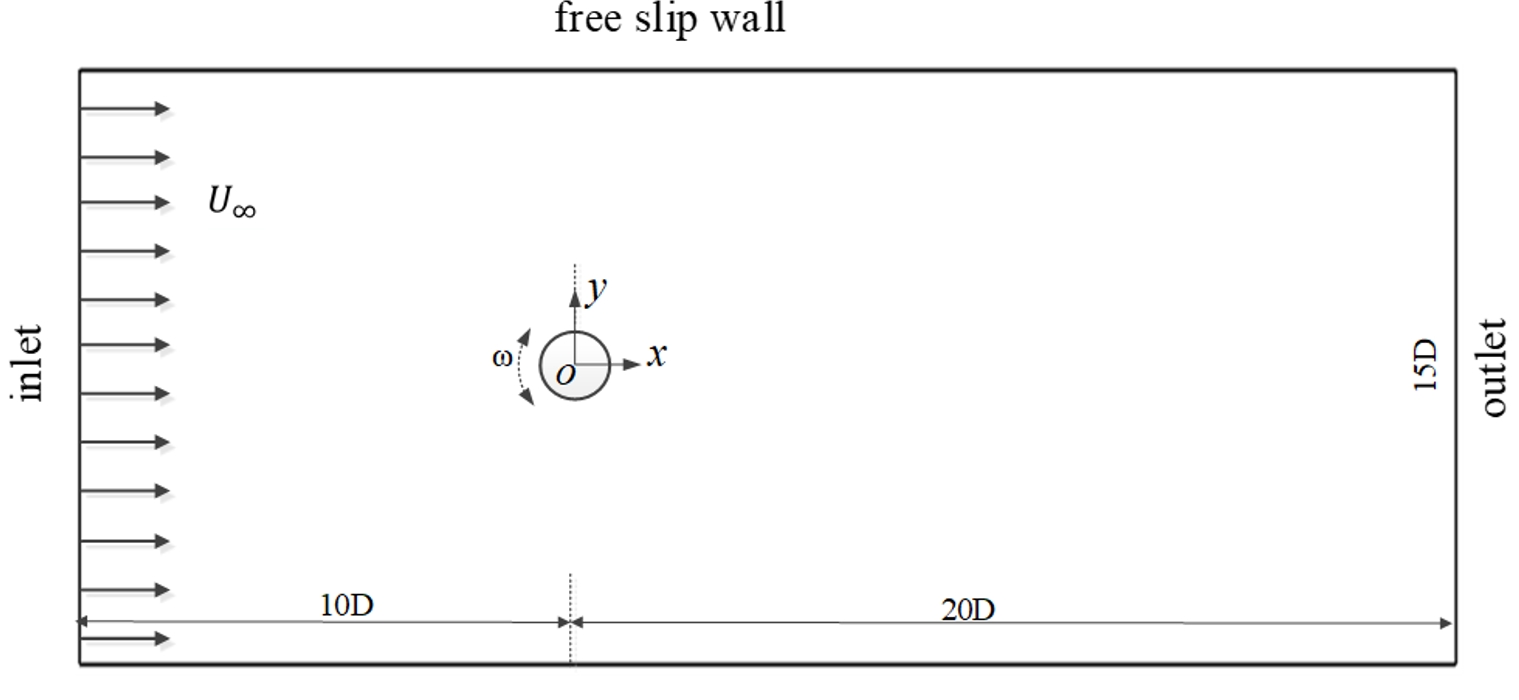}}
  \caption{Schematic representation of the setup and computational domain}
\label{fig:k4}
\end{figure}
\\As one of the pioneering modal analysis methods, proper orthogonal decomposition has been applied in various fields  \citep{riches2018proper}. It is mostly used to discover the hidden deterministic structures of the flow field and to realize the flow field reconstruction \citep{chang2021reduced}. Firstly, fully developed flow field data was collected. Here, the number of data samples covered 20 complete vortex-shedding cycles. The time interval between each sample was $T/20$, i.e., there were 20 samples per vortex-shedding period. The time-varying flow field can be decomposed into a finite sum of spatial eigenfunctions (coherent POD mode) multiplied by a temporal modal coefficient. The POD modes include mean mode (M0) and dynamic modes (M1, M2, …). 
\\The flow field include three components, i.e., streamwise velocity, vertical velocity and pressure field. We take the mean streamwise velocity as an example. Firstly, we investigate the performance of Grassmannian-kNN Learning for M0 (i.e., mean mode). Here, we only use $N$ = 120 training samples $X_{\mathrm{i}}\in\mathbf{R}^{\mathrm{N}\times2},$ with corresponding model outputs $Y_{\mathrm{i}}\in\mathbf{R}^{\mathrm{N}\times12600},$ where the square computational domain has been discretized in $w$ = 126 × 100 mesh points. GDMaps converged to a Grassmann manifold dimension of $p$ = 12 which results in matrices on the Grassmannian $\{U_i,V_i\in G(12,100)\}^{\mathrm{N}}_{\mathrm{i=1}}$ for the training dataset, based on the residuals computed by the eigen decomposition of the Markov matrix. The first $q$ = 4 non-trivial diffusion coordinates are considered, specifically $\{\theta_1,\theta_2,\theta_3,\theta_4\}$ to represent the embedding structure for the datasets with $N$ = 120. The diffusion coordinates for Grassmann manifold dimension of mean flow field are shown in Figure \ref{fig:k5}. In Figure \ref{fig:k6}, we present the first forth Karcher means flow field for different clusters. The Grassmannian-KNN Learning can accurately predict the mean flow field shown in Figure \ref{fig:k10} (a, b and c). they share similar features in space, but with different wake length.
\\At $Re$ = 100, the cylinder experienced strong unsteady vortex shedding. The coherence mode is of importance in reconstructing the flow field. POD mode energy rank for baseline case and the self-rotation are shown in Figure \ref{fig:k7}. Apparently, the first two modes dominate the wake, which are conjugate.  Thus, we only investigate the performance of Grassmannian-kNN Learning for M1. Note that, the training samples is set as $N$ = 256, which are larger than the mean mode case, due to the reason that the dynamic modes change greater than the mean mode case. The other setups are same the case for M0. The diffusion coordinates for Grassmann manifold dimension of the dynamic mode M1 are shown in Figure \ref{fig:k8}, here,  the Grassman manifold describes all modes with 12 latent variables. Three clusters are enough to classify the physical flow pattern. Karcher means dynamic modes at different cluster are plot in Figure \ref{fig:k9}. Cluster K3 in Figure \ref{fig:k9} (d) denotes the cases which show classical vortex shedding, i.e., Karman vortex shedding. The flow dynamics seems have little difference with that of the uncontrolled case. In Cluster K1 shown in Figure \ref{fig:k9} (b), two antisymmetric pattern stay near wake, while several antisymmetric small-scale structures appear in far wake. For  Cluster K2 shown in Figure \ref{fig:k9} (c),  two distinct antisymmetric pattern attach in the trailing of the cylinder, and no large-scale coherence structures appear in near and far wake. This means the vortex shedding is completely suppressed by self-rotation, and only shear layers show oscillation (like K-H vortices). The Grassmann-KNN Learning can accurately predict the dynamics mode shown in Figure \ref{fig:k10} (b).
\begin{figure}
  \centerline{\includegraphics[scale=0.2,angle=0]{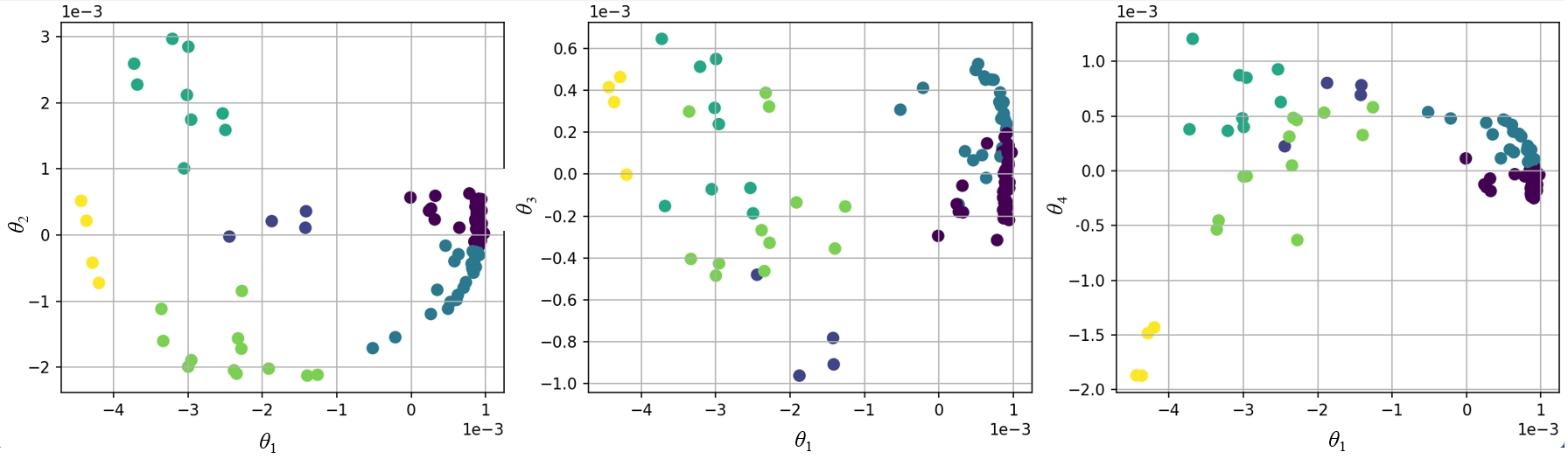}}
  \caption{The diffusion coordinates for Grassmann manifold dimension of mean flow field. }
\label{fig:k5}
\end{figure}
\begin{figure}
  \centerline{\includegraphics[scale=0.3,angle=0]{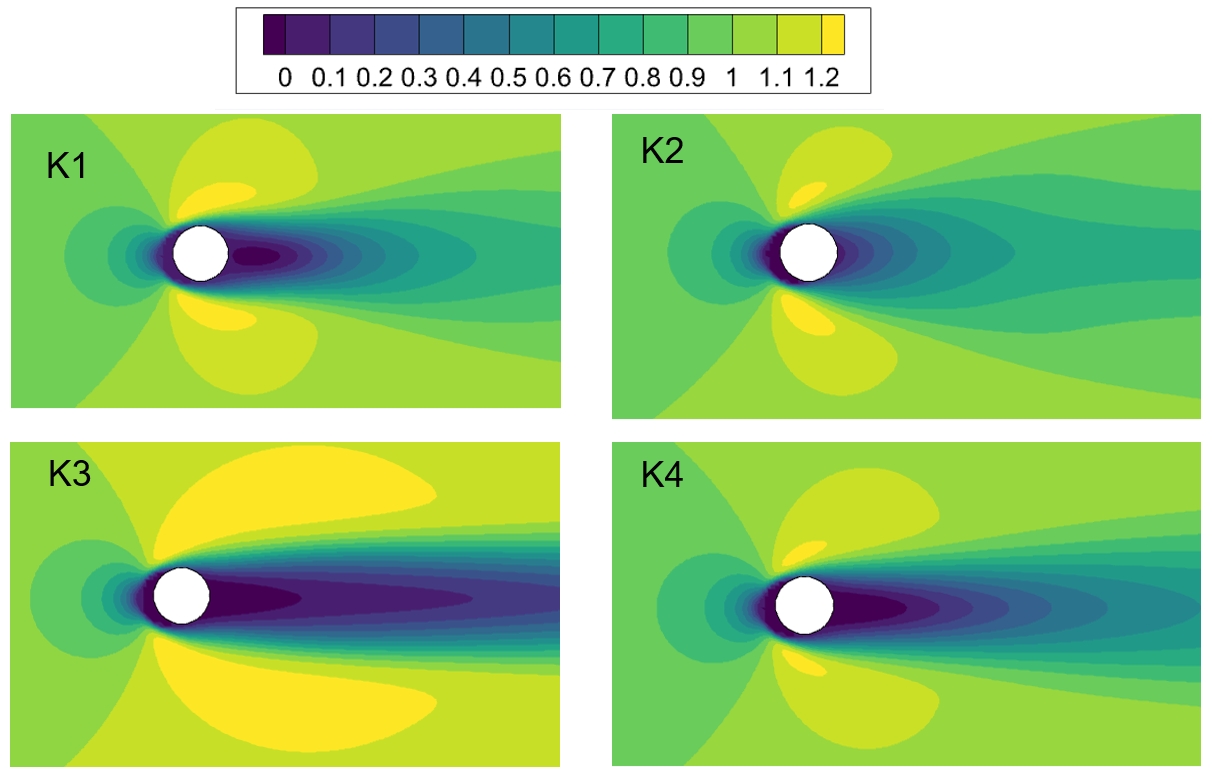}}
  \caption{Karcher means flow field (M0) for 4 different clusters. }
\label{fig:k6}
\end{figure}

\begin{figure}
  \centerline{\includegraphics[scale=0.25,angle=0]{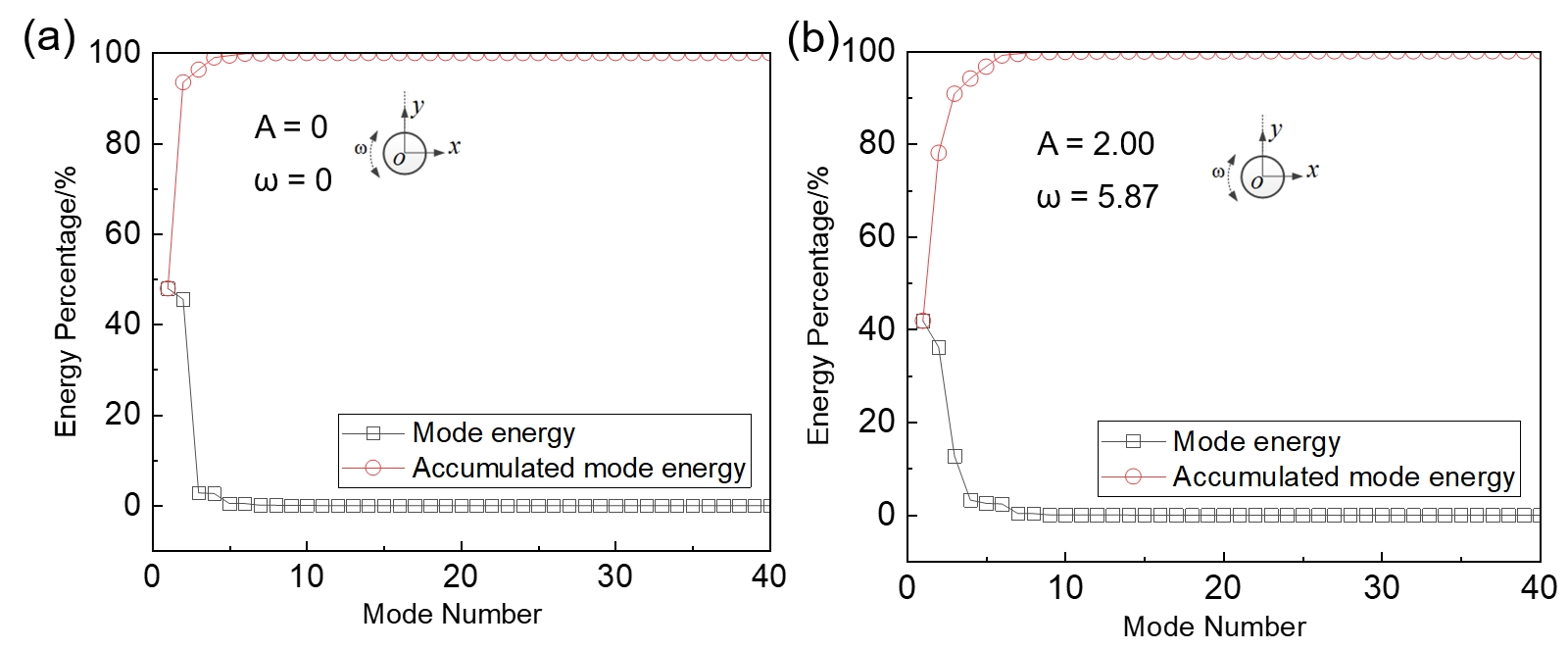}}
  \caption{Illustration of the present motivation and objectives}
\label{fig:k7}
\end{figure}
 \begin{figure}
  \centerline{\includegraphics[scale=0.25,angle=0]{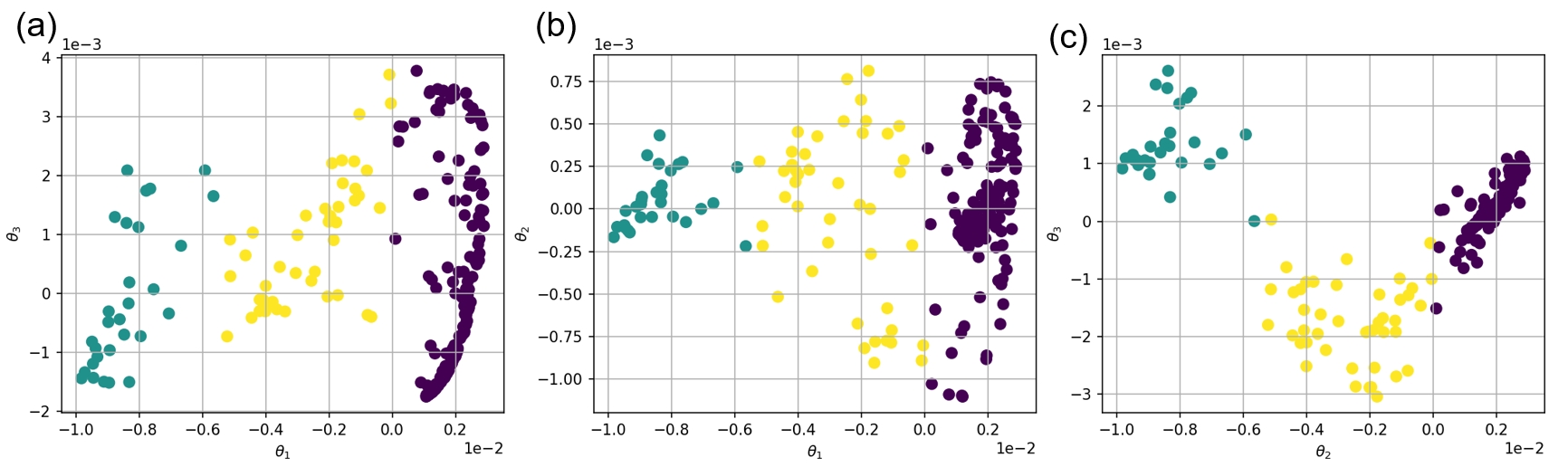}}
  \caption{Coherence mode at different cluster represented by the first three non-trivial diffusion coordinates. }
\label{fig:k8}
\end{figure}
\begin{figure}
  \centerline{\includegraphics[scale=0.25,angle=0]{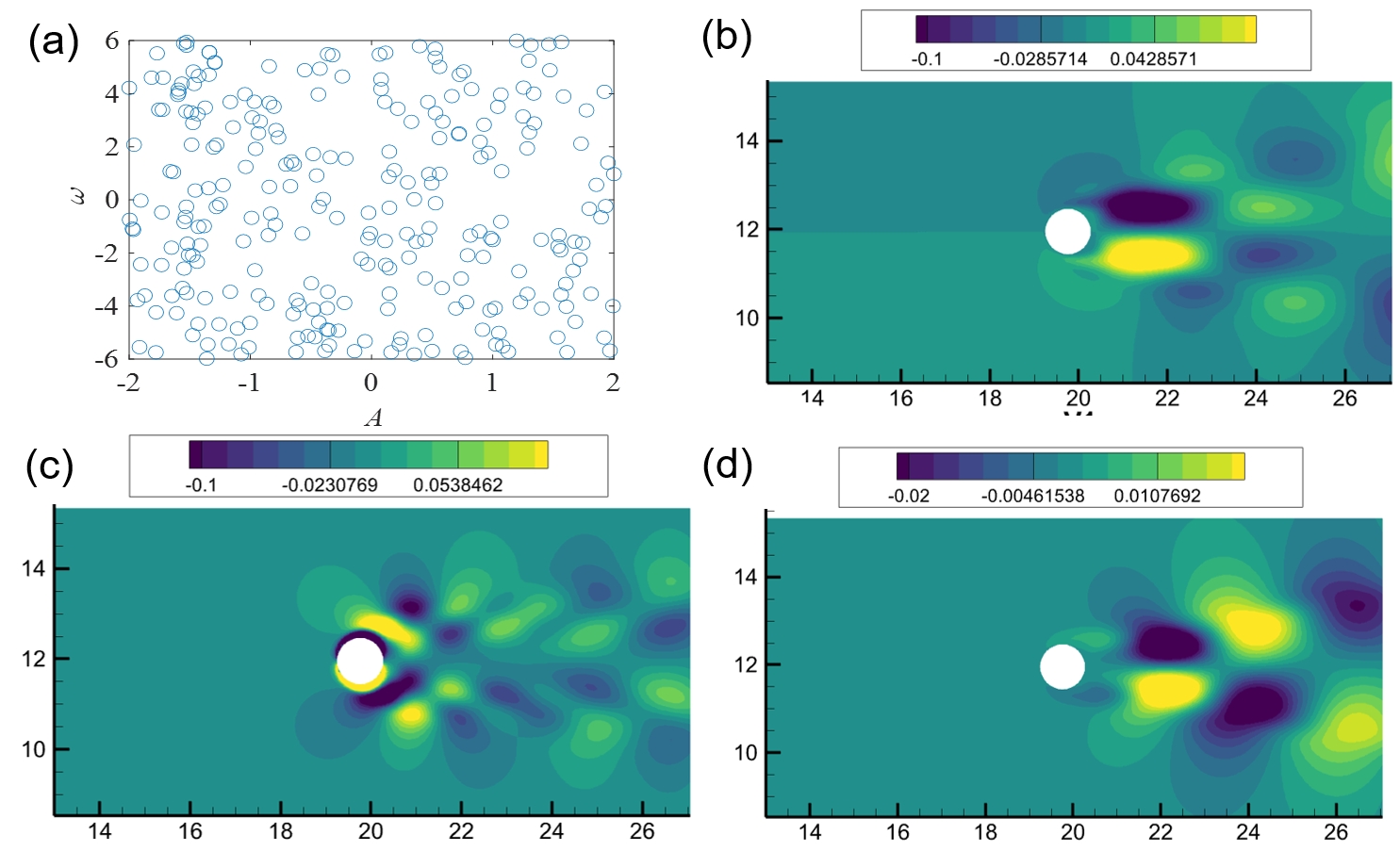}}
  \caption{Karcher means coherence mode at different cluster represented by the first three non-trivial diffusion coordinates.}
\label{fig:k9}
\end{figure}
\begin{figure}
  \centerline{\includegraphics[scale=0.3,angle=0]{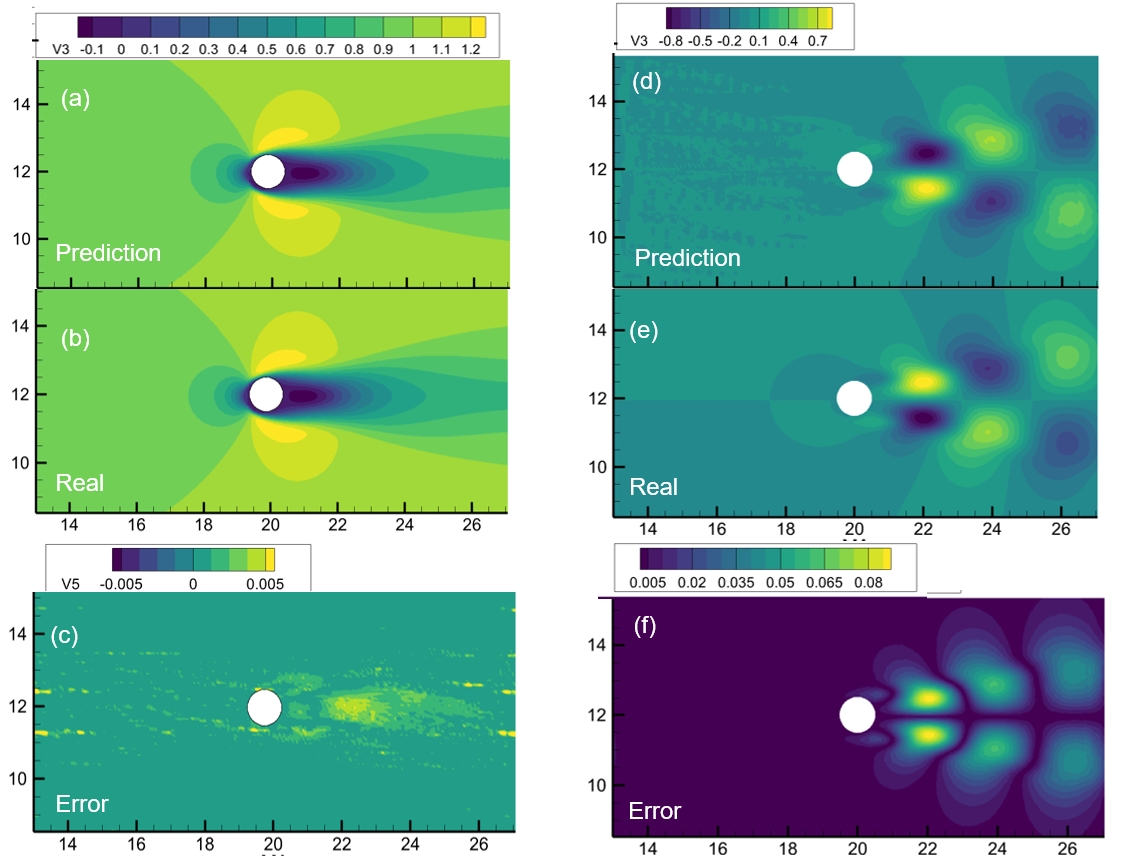}}
  \caption{Illustration of the present motivation and objectives}
\label{fig:k10}
\end{figure}
\begin{table}
  \begin{center}
\def~{\hphantom{0}}
  \begin{tabular}{lc}
      Parameters& Values\\[3pt]
       $Re$& 100\\
       $A/D$& Uniform [-2.0~2.0 ]\\
       $\omega$& Uniform [-6.0~6.0]\\
  \end{tabular}
  \caption{Test conditions of the cylinder}
  \label{tab:k2}
  \end{center}
\end{table}
\section{Conclusion}\label{sec:Figures_Tables}
In this study, we address a challenge of reduced-order modeling: accurate low-dimensional representations for a large range of operating conditions. The dynamic snapshot output as well as parameterized space are compressed in a Grassman manifold using PCE as encoder and $\text{kNN}$ as decoder. The methodology is tested for a simple dielectric cylinder in homogeneous electric field firstly. We used $N$ = 400 training samples to predict the electric field via the proposed model. It shows fast and accurate performance.
\\The cylinder oscillating rotations amplitude and frequency are then used as two independent control parameters to test the model on predicting dynamic coherence modes. Starting point of the reduced-order modeling is a data set with 20 statistically representative post-transient snapshots at $Re=100$ for 216 sets of cylinder oscillating rotations covering uniformly a box with circumferential velocities between $\text{-2}$ and 2 and frequency between $\text{-6}$ and 6. The Grassman manifold describes all snapshots with 3 latent variables and few percent representation error. Intriguingly, all latent variables are aligned with clear physical meanings. It is possible to estimate the full flow state with small reconstruction errors just by knowing the actuation parameters.

\appendix

\section{}\label{appA}

\subsection {Grassman tangent space and Grassmannian distance}
The Grassmann manifold is a smooth and continuously differentiable manifold, which enables numerous mathematical operations such as differentiation and optimization. Given that the Grassmann manifold is smooth and continuously differentiable, one can define a trajectory $\gamma$(z), $z$ $\in $ [0, 1], known as geodesic, defining the shortest path between two points, $\Psi$(0) = $F_{0}$ = span($\gamma_{0}$) and $\gamma$(1) = $F_{1}$ = span($\Psi_{1}$), on the manifold $\mathcal{G}(p,n)$. The derivative of this trajectory at any point $F$ (represented by $\Psi$) defines the tangent space $(\mathcal{T}_{\mathcal{F}}\in(p,n)),$ which is given by the set of all tangent vectors $\mathcal{T}$ such that $\mathcal{T}(p,n)=\{\boldsymbol{T}\in\mathbb{R}^{n\times p}:\boldsymbol{T}^{\intercal}\boldsymbol{\Psi}=0\}$. Therefore, given a tangent space $\mathcal{T}(p,n)$ at $F_{0}$, one can map $\mathcal{T}_{1}$ onto the Grassmannian point $\gamma$(1) = $F_{1}$ represented by $\Psi_{1}$ via the exponential map \citep{zhang2018grassmannian}

\begin{equation}\mathbf{\Psi}_1=\exp_{\mathbf{X}_0}(\mathbf{T}_1)=\exp_{\mathbf{X}_0}(\mathbf{U}\mathbf{S}\mathbf{V}^T)=\mathbf{\Psi}_0\mathbf{V}\mathrm{cos}(\mathbf{S})\mathbf{Q}^T+\mathbf{U}\mathrm{sin}(\mathbf{S})\mathbf{Q}^T\label{A1}\end{equation}
where $\mathbf{Q}\in\mathbb{R}^{n\times n}$ is an orthogonal matrix satisfying the following expressions.
\begin{equation}\mathbf{V}\mathrm{cos}(\mathbf{S})\mathbf{Q}^T=\mathbf{\Psi}_0^T\mathbf{\Psi}_1\label{A2}\end{equation}
\begin{equation}\mathbf{U}\mathrm{sin}(\mathbf{S})\mathbf{Q}^T=\mathbf{\Psi}_1-\mathbf{\Psi}_0\mathbf{\Psi}_0^T\mathbf{\Psi}_1\label{A3}\end{equation}
After appropriate manipulation, one can obtain the following expression
\begin{equation}\mathbf{U}\mathrm{tan}(\mathbf{S})\mathbf{V}^T=(\mathbf{\Psi}_1-\mathbf{\Psi}_0\mathbf{\Psi}_0^T\mathbf{\Psi}_1)(\mathbf{\Psi}_0^T\mathbf{\Psi}_1)^{-1}\label{A4}\end{equation}
Consequently, one can write the logarithmic map from the Grassmannian to the tangent space:
\begin{equation}\log_X(\Psi_1)=\mathbf{U}\tan^{-1}(\mathbf{S})\mathbf{V}^T\label{A5}\end{equation}
The Grassmann manifold's properties also give a natural notion of distance between points on the Grassmann manifold. Not only there are actually several meanings of distance that level out to the differences in principal angels between subspaces. It is easy to see that the cosine of the principal angles $\alpha_1\in[0,\pi/2]$ between a pair of subspaces $X=\mathrm{span}(\Psi_{\mathrm{x}})$ and $Y=\mathrm{span}(\Psi_{\mathrm{y}})$  can be computed from the singular values of $\mathbf{\Psi}^{\mathrm{T}}_{\mathrm{x}}\mathbf{\Psi}_{\mathrm{y}}=\bar{\mathrm{U}}\bar{\mathrm{S}}\bar{\mathrm{V}}^{\mathrm{T}},$  where $\bar{\mathrm{U}}$ $\in\mathrm{O}(k),\bar{\mathrm{V}}\in\mathrm{O}(l),\mathrm{~and~}\bar{\mathrm{S}}=\mathrm{diag}(\sigma_1,\sigma_2,\ldots,\sigma_p).$So, we have $\alpha_{i}=\cos(\sigma_{i}).$ 
 The geodesic distance, $d_{\mathcal{G}(p,n)}(X,Y),$ shows the distance along the geodesic curve $\gamma(\mathrm{z})$ parameterized by $z$ $\in $ [0, 1], and is calculated by $d_{\mathcal{G}(p,n)}(X,Y)=\|\mathbf{s}\|_{2},$ where $s$ = ($\alpha_{1},\alpha_{2},\ldots,\alpha_{p}$ ) is the principal angles. 

Kernel-based dimensionality reduction methods rely on defining a proper kernel $k(x_{i},x_{j})\mathrm{with}\sum c_{i}c_{j}(x_{i},x_{j})\geq0,$where $c_{i},c_{j}\in\mathrm{R}.$ The most commonly used $k(X_i,Y_j)=\exp(-||X_i-Y_j||/\epsilon),$  where  $X_{i,}Y_{j}$ are the high dimensional data. and $\epsilon$ is a length-scale parameter. Nevertheless, the Gaussian kernel is inappropriate to describe the manifolds of data. Grassmannian kernels have the advantage of being equipped with this property, contrary to standard discriminative models when analyzing high-dimensional data.
This property is achieved by Grassmannian kernels, which is specially useful in high-dimensional data analysis. One of the most well-known kernels over the Grassmannian is the real symmetric map $\mathcal{G}(p,n)\times\mathcal{G}(p,n)\to\mathbb{R}$ in form of a kernel, which is also referred to as Grassmannian kernel defining an embedding into a reproducing kernel Hilbert space. Additionally, since a Grassmannian kernel is invariant to basis choice and is also positive semi-definite. Several families of other Grassmannian kernels are also defined in the literatures \citep{zhang2018grassmannian,ye2022optimization}. Wherein, the widely used kernels are the Binet–Cauchy and projection kernels. The Binet–Cauchy kernel is built via taking the manifold $\mathcal{G}(p,n)$ into a projective space. For instance, there are two subspaces $X=\mathrm{span}(\mathbf{\Psi}\mathbf{x})$ and $Y=\mathrm{span}(\mathbf{\Psi}\mathbf{y})$, the kernel is defined as
\begin{equation}k_{bc}(X,y)=\det\left(\Psi_x^T\Psi_y\right)^2\label{A6}\end{equation}
or 
\begin{equation}k_{bc}(X,Y)=\prod_{i=1}^p\cos^2(\theta_i)\label{A7}\end{equation}
Another kernel widely used kernel is the projection kernel. It is defined as
\begin{equation}k_{pr}(X,Y)=\parallel\Psi_{x}^{T}\Psi_{y}\parallel_{F}^{2}\label{A8}\end{equation}
Or 
\begin{equation}k_{pr}(\mathcal{X},\mathcal{Y})=\sum_{i=1}^p\cos^2\left(\theta_i\right)\label{A9}\end{equation}
More details on Grassmannian kernels can be found in \cite{dos2022grassmannian}.
\subsection {Grassmannian Diffusion Maps}
Diffusion maps (DMaps) belong to the family of manifold learning approaches, it is underpinned by approximating a Markov transition probability matrix in association with a random walk on graph connecting data \citep{kontolati2022manifold}. The graph (or network) has nodes corresponding to data points and edges indicating the connections between these data points which are weighted by transition probabilities that capture local similarities of pair-wise points. The graph structure can be parameterized by the diffusion coordinates which represent the underlying low-dimensional manifold (embedding) of the data. This special parameterization is identified by the way that we choose the eigenvectors of the Markov matrix. Builds a graph linking subspaces on the Grassmann manifold \citep{coifman2006diffusion}. In this part, we briefly describe the fundamental concepts of GDMaps which are adopted from \citep{kontolati2022manifold}.
Consider a set of points (projected high-dimensional data) on the Grassmann manifold $\mathcal{G}(p,n)$ given by $\mathcal{G}_N=\{y_1,\ldots,y_N\}$ and a positive semi-definite Grassmamian kernel k:$\mathcal{G}_{p,n}\times\mathcal{G}_{p,n}\to\mathbb{R}$,  also known as the diffusion kernel. If we consider a random walk over $\mathcal{G}_N$  having probability distribution $f$, $W_N=(\mathcal{G}_N,f,\mathbf{P}),$ we can construct the transition probability matrix $P$ as follows. First, we construct the degree matrix

Let a Grassmamian Kernel $k$: $\mathcal{G}_{p,n}\times\mathcal{G}_{p,n}\to\mathbb{R}$ give the similarities (potential propagations) between elements of projected high-dimensional data $\mathcal{G}_N=\{y_1,\ldots,y_N\}$ on the Grassmann manifold $\mathcal\{G\}(p,n)$ (also known as diffusion kernel). Let's start with a random walk on $\mathcal{G}_N$ with distribution $f$, $W_N=(\mathcal{G}_N,f,\mathbf{P}),$ the associated transition probability matrix $P$ is constructed as following. Step 1: Building the degree matrix
\begin{equation}D_{ii}=\sum_{j=1}^Nk(y_1,y_j)\label{A10}\end{equation}
where $D_{ii}$ is a diagonal matrix $D\in\mathbb{R}^{N\times N}$.
Next, the kernel is normalized as 
\begin{equation}\kappa_{ij}=\frac{k_{ij}}{\sqrt{D_{ii}D_{jj}}}\label{A11}\end{equation}
and the transition probability matrix $P_{ij}$ of the random walk over the Grassmannian is given by
\begin{equation}P_{ij}^t=\frac{\kappa_{ij}}{\sum_{k=1}^ND_{ik}}\label{A12}\end{equation}
Propagating the Markov chain is the same as running a diffusion process on the manifold: this allows us to recover from data points some geometric structure of what resides naturally on the first $q$ Grassmannian $\left\{\xi_k\right\}_{k=1}^q$, with $\xi_k\in$ $\mathbb{R}^N$ and corresponding eigenvalues $\left\{\lambda_k\right\}_{k=1}^q.$ Hence, the diffusion coordinates are defined as
\begin{equation}\Theta_j=(\theta_{j0},...,\theta_{jq})=(\lambda_0\xi_{j0},...,\lambda_q\xi_{jq})\label{A13}\end{equation}
where~$\mathrm\xi_{jk}$  is the position$j$ of~ $\xi_k$. Due to the spectral decay of the eigenvalues of the sparse Markov matrix, usually a small $q$ is sufficient to capture the essential geometric structure of the dataset. 
We look back to two very basic and defining features of GDMaps as opposed to custom DMaps: (1) Data points lie on : the data space we perform DMaps in, is actually subspaces that collectively span all possible subspaces in which the original data exists. (2) Grassmannian kernel used: The Grassmannian kernel is a powerful kernel that measures the similarity between subspaces. The reason for utilizing a subspace representation is mainly the inadequacy of measuring similarity among very high dimensional objects, further discussed in \citep{kontolati2022manifold}.
\subsection {Polynomial Chaos Expansion (PCE)}

We consider a model with $k$-variate random variables being denoted as $\mathcal{M}(\mathbf{X}), \mathbf{X}$ defined on $(\Omega,\Sigma,P)$ and characterized by the joint probability density function (PDF) $\varrho_X:Z\to\mathbb{R}_{\geq 0}$ where $Z \in\mathbb{R}^k$ is the image space, $Ω$ is the sample space, $\Sigma$ is the set of events and $P$ the probability measure. Herein $X$ is considered as a set of independent random variables and achieved results are consequently relying on this assumption the rest of work, but it should be noted that PCE method can also be used for dependent random variables \citep{blatman2011adaptive,jakeman2015enhancing,loukrezis2020robust,hadigol2018least,conrad2013adaptive}. 
Next, having that the model $\mathcal{M}$ (in case it is a valid using the notation of Doob-Dynkin lemma \citep{bobrowski2005functional}) so that output of that seems like another random variable as function on input data $\mathbf{X}$. For simplicity here, we will just consider a single model output with $Y(\omega)=\mathcal{M}(\mathbf{X}(\omega))\in\mathbb{R}, \quad \omega\in\Omega.$ However, extending the previous setup to multivariate outputs is relatively easy as we can simply apply the PCE approximation described in the next few lines element-wise. In the following, we will use the same notation to represent a random variable $\mathbf{X}$ but the specific usage should be clear from context.
Under the assumption of a single model output, the PCE is a spectral approximation of the form
\begin{equation}\mathcal{M}(\mathbf{X})\approx\widetilde{\mathcal{M}}(\mathbf{X})=\sum_{\mathrm{s}=1}^\mathrm{S}\mathrm{c}_\mathrm{s}\Xi_\mathrm{s}\left(\mathbf{X}\right)\label{A14}\end{equation}
where $c_\mathrm{s}$ are scalar coefficients and $\Xi_\mathrm{s}$ are multivariate polynomials that are orthonormal with respect to the joint PDF $\varrho_{X}$, such that
\begin{equation}\mathbb{E}[\Xi_s\Xi_t]=\int_Z\Xi_s(\mathbf{X})\Xi_t(\mathbf{X})\varrho_X(\mathbf{X})\mathrm{d}\mathbf{X}=\delta_{\mathrm{st}}\label{A15}\end{equation}
where $\delta_{\mathrm{st}}$ denotes the Kronecker delta. Depending on the PDF $\varrho_X,$ the orthonormal polynomials can be selected under the Wiener-Askey scheme, or from a numerical construction. Note that given $X=X_{1},\ldots,X_{k}$ are independent random variables, the joint PDF is

\begin{equation}\varrho_X(\mathbf{X})=\prod_{i=1}^k\varrho_{x_i}(X_i)\label{eqA16}\end{equation}

with $\varrho_{x_i},$ represents the marginal PDF of random variable $X_{i}$. Therefore, the multivariate orthogonal polynomials are built as 

\begin{equation}\Xi_s(\mathbf{X})\equiv \prod_{i=1}^k\xi_i^{s_i}(X_i)\label{A16}\end{equation}

where $\xi_i^{s_i}$ are univariate polynomials of degree $s_i\in\mathbb{Z}_{\geq0}$ that are orthonormal with respect to the univariate PDF $\varrho_x$, so that

\begin{equation}\mathbb{E}[\xi_{i}^{s_{i}}\xi_{i}^{t_{i}}]=\int_Z \xi_i(X_i)^{s_i}\varrho_ {x_i}(X_i)dx= \delta_{s_it_i}.\label{A16}\end{equation}

A multivariate polynomial of degree $\mathbf{s}$ corresponds to a multi-index $\mathbf{s}=(s_{1},\ldots,s_{k})$, that is uniquely associated with the single index $s$ and now should be parametrized as follows

\begin{equation}\mathcal{M}(\mathbf{X})\approx\tilde{\mathcal{M}}(\mathbf{X})=\sum_{\mathbf{s}\in\Lambda}\mathfrak{c}_\mathbf{s}\Xi_\mathbf{s}(\mathbf{X})\label{A17}\end{equation}

where $\Lambda$ is a multi-index set and with cardinality $\#\Lambda=S.$ The multi-index set, \(\Lambda\), plays the central role in constructing the PCE: it defines which polynomials (and hence their associated coefficients) form the PCE. The most popular option, including the one used in this paper, is a total-degree multi-index set, where $\Lambda$ contains all multi-indices for which $\|s\|_1\leq s_{max},s_{max}\in\mathbb{Z}_{\geq0}. $ Then the dimension of PCE $S=\frac{(s_{max}+k)!}{s_{max}!k!},$  where the complexity here would scale polynomially with the input dimension $k$ and the  maximum degree ${s}_{max}$.   Due to curse of dimensionality, the list generated by our procedure could be large in high-dimensional input random variable $\mathbf{X}$, Therefore several sparse PCE algorithms were developed in the literature for construction of $\Lambda$, to reduce this impact,see \citep{blatman2011adaptive,jakeman2015enhancing,loukrezis2020robust}.
Given that the multi-index set $\Lambda$ is chosen, all that remains to fully specify a PCE is to determine the coefficients. There are several methods in the literature for computing the PCE coefficients. The latter choice is used in this work as well, and the PCE coefficients are computed by solving the penalized least squares problem

\begin{equation}arg\min_{c} \left( \frac{1}{N}\sum_ {i=1}^N (f(X_i)-\ sum_ s c_s\xi_s(X_i))^2+reg(J(c)) \right) \label{A18},\end{equation}

where $\lambda\in\mathbb{R}$ is a penalty factor, $J(c)$ a penalty function acting on the vector of PCE coefficients $c\in\mathbb{R}^{\#\Lambda}$, 
and $\mathcal{X}=\{X_i\}_{i=1}^{\mathcal{N}}$ an experimental design (ED) of random variable realizations with corresponding model outputs $\mathcal{Y}=\{Y_i\}_{i=1}^{\mathcal{N}}$. There are many possible choices for the penalty function $J(c)$: e.g. the For both least absolute shrinkage and selection operator and ridge regression we must solve Problem. If you take out the penalty term, it leads us to an ordinary least squares (OLS) regression problem.


\bibliographystyle{jfm}
\bibliography{jfm}


\end{document}